%% file: main.tex
\documentclass[11pt]{article}
\usepackage[utf8]{inputenc}
\usepackage[a4paper, width=160mm, top=30mm, bottom=40mm]{geometry}
\usepackage{indentfirst}
\usepackage{graphicx}

\usepackage{amsmath}
\usepackage{amsfonts}
\usepackage{amssymb}
\usepackage{amsthm}
\usepackage{titling}
\usepackage{enumerate}
\usepackage{verbatim} 
\usepackage{authblk}

\theoremstyle{definition}

\newtheorem{remark}{Remark}
\usepackage{titlesec}
\DeclareUnicodeCharacter{2061}{}

\usepackage{newpxtext}
\usepackage{newpxmath}

\usepackage{array}
\usepackage{multirow}

\newcolumntype{C}[1]{>{\centering\arraybackslash}p{#1}}

\usepackage{tikz}
\usetikzlibrary{shapes,arrows}
\usetikzlibrary{intersections}

\titleformat{\paragraph}
{\normalfont\normalsize\bfseries}{\theparagraph}{1em}{}
\titlespacing*{\paragraph}
{0pt}{3.25ex plus 1ex minus .2ex}{1.5ex plus .2ex}

\usepackage[style=authoryear, bibencoding=auto, hyperref=true, maxcitenames=2, maxbibnames = 15, sorting = nyt, uniquelist=false]{biblatex}
\DeclareNameAlias{sortname}{last-first}

\usepackage[hyperfootnotes=true]{hyperref}
\hypersetup{colorlinks, citecolor=blue}

\DefineBibliographyStrings{german}{%
  andothers = {et al.},
}

\DeclareCiteCommand{\textcite}
 {\boolfalse{cbx:parens}}
  {\usebibmacro{citeindex}%
    \printtext[bibhyperref]{\printnames{labelname}%
      \printtext{ (\printfield{year}\printtext{)}}}}
 {\ifbool{cbx:parens}
  {\bibcloseparen\global\boolfalse{cbx:parens}}
  {}%
 \multicitedelim}
{\usebibmacro{textcite:postnote}}
  
\DeclareCiteCommand{\parencite}[\mkbibparens]
  {\usebibmacro{prenote}}
  {\usebibmacro{citeindex}%
    \printtext[bibhyperref]{\usebibmacro{cite}}}
  {\multicitedelim}
  {\usebibmacro{postnote}}
  
\DeclareCiteCommand{\cite}
  {\usebibmacro{prenote}}
  {\usebibmacro{citeindex}%
   \printtext[bibhyperref]{\usebibmacro{cite}}}
  {\multicitedelim}
  {\usebibmacro{postnote}}

\usepackage{xpatch}

\begin{document}

\begin{titlingpage}
    \centering
    \vfill
     \LARGE{
        Forecasting High-Dimensional Covariance Matrices of Asset Returns with Hybrid GARCH-LSTMs}\par
        \vskip1cm
       \Large
        {Lucien Boulet \footnote{\noindent lucien.boulet@gmail.com}\\
        \large
        MSc Student, Financial Markets,\\\vspace{-1mm}
        Université Paris-Dauphine - PSL.\\
        MSc Student, Pure Mathematics,\\\vspace{-2.5mm}
        Sorbonne Université.}\\
        \vspace{5mm}
        \large
        August 2021\\
      \vskip1cm
    \begin{abstract}
        \input{Abstract/Abstract}

    \end{abstract}
\end{titlingpage}

\section{Introduction}
\input{Introduction/Introduction}

\section{Minimum Variance Portfolio}
\label{sec:2}
\input{MVP/MVP}

\section{Univariate and Multivariate GARCH models}
\label{sec:3}

\subsection{The GARCH model}
\input{Econometric_models/GARCH}

\subsection{Dynamic Conditional Correlation GARCH}
\input{Econometric_models/DCC}



\subsubsection{Estimating DCC in Large Dimensions}
\input{Econometric_models/Robustifying_multivariate_garch}

\section{Deep Learning Models}
\label{sec:4}
\input{Deep_Learning/Intro_section}

\subsection{Deep Feedforward Neural Networks}
\input{Deep_Learning/FNN}

\subsection{Recurrent Neural Networks}
\label{sec:3.1.2}
\input{Deep_Learning/RNNs}

\subsection{Long Short-Term Memory Neural Networks}
\input{Deep_Learning/LSTM}




\section{A New Hybrid Multivariate GARCH-LSTM Model}
\label{sec:5}
\subsection{The Rational Behind Hybrid Models}
\input{Hybrid_model/Intro_section_hybrid}

\subsection{The GARCH-LSTM-DCC Model}
\input{Hybrid_model/The_GARCH-LSTM-DCC}

\subsection{Advantages and Limits of the GARCH-LSTM-DCC Model}
\label{sec:5.3}
\input{Hybrid_model/Advantages_and_limits}

\section{Empirical Application}
\label{sec:6}
\input{Empirical_part/Intro_section_empirical}

\subsection{Dataset and Experimental Protocol}

\subsubsection{Dataset}
\input{Empirical_part/Data}

\subsubsection{Experimental Protocol}
\label{sec:6.1.2}
\input{Empirical_part/Economic_Evaluation}

\subsection{Results}
\input{Empirical_part/Results}

\section{Conclusion}
\input{Conclusion/Conclusion}

\section{Acknowledgments}
\input{Acknowledgement/ack}

\newpage
\appendix

\section{Numerical Details}
\label{app:Numerical Details}
\input{Appendices/Numerical_details}

\newpage
\nocite{*}
\printbibliography{}

\end{document}

%% file: Abstract/Abstract.tex
Several academics have studied the ability of hybrid models mixing univariate Generalized Autoregressive Conditional Heteroskedasticity (GARCH) models and neural networks to deliver better volatility predictions than purely econometric models. Despite presenting very promising results, the generalization of such models to the multivariate case has yet to be studied. Moreover, very few papers have examined the ability of neural networks to predict the covariance matrix of asset returns, and all use a rather small number of assets, thus not addressing what is known as the curse of dimensionality. The goal of this paper is to investigate the ability of hybrid models, mixing GARCH processes and neural networks, to forecast covariance matrices of asset returns. To do so, we propose a new model, based on multivariate GARCHs that decompose volatility and correlation predictions. The volatilities are here forecast using hybrid neural networks while correlations follow a traditional econometric process.
After implementing the models in a minimum variance portfolio framework, our results are as follows. First, the addition of GARCH parameters as inputs is beneficial to the model proposed. Second, the use of one-hot-encoding to help the neural network differentiate between each stock improves the performance. Third, the new model proposed is very promising as it not only outperforms the equally weighted portfolio, but also by a significant margin its econometric counterpart that uses univariate GARCHs to predict the volatilities.

%% file: Introduction/Introduction.tex
For almost forty years now, since the introduction of the Autoregressive Conditional Heteroskedasticity (ARCH) model by \textcite{arch}, forecasting asset return volatility has been the cornerstone of financial econometrics. Indeed, volatility, which generally refers to the amplitude of returns’ fluctuations over a given period of time, is a key measure of risk used in many financial applications.
In many financial application that involve several assets, it is often as important to be able to predict to what extent these assets interact with each other, in addition to predict their volatility.
For example, since the modern portfolio theory of \textcite{Markowitz}, being able to produce reliable forecasts of covariances and correlations between asset returns has become a key of building performing asset portfolios. 

For these reasons, numerous models have been introduced by academics and market practitioners to give robust predictions of the conditional covariance matrix. Among these, Generalized ARCH (GARCH)-type models seem to be particularly popular both for volatility and covariance predictions. While the reliability of GARCH models have been shown multiple times in a unidimensional setting, their extension to a high-dimensional framework presents some difficulty. The main difficulty is what is often called the 'curse of dimensionality'. This term refers to the fact that in high dimensions, such model face a trade-off between flexibility and computational feasibility. Indeed, a rigorous specification of the one-step ahead conditional covariance matrix of asset returns would require all its elements to be dependant on all the elements of the conditional covariance matrix at the preceding time step, or at least on their own past values. However, such parametrizations result in an explosion in the number of parameters when the number of assets under consideration increases. As such, they can only be implemented in their original form for typically less than five assets, and have to be drastically simplified when large-dimensional matrices are considered.

To circumvent this problem, a specific class of multivariate GARCHs was introduced by \textcite{CCC} with the Constant Conditional Correlation (CCC) model and further extended by \textcite{DCC} with the Dynamic Conditional Correlation (DCC) model. The idea behind these is to model separately on one side conditional variances using univariate GARCHs, and on the other side conditional correlations as constants (CCC) or using a multivariate GARCH (DCC). Rendered robust for large dimensions by \textcite{Engle2017}, the DCC (and the corrected DCC (cDCC) of \textcite{cDCC}) seem to be the most precise models usable when the number of asset is large.

In the univariate case, 
several academics have tested the ability of artificial neural networks (ANNs) to produce better volatility predictions than econometric models. This interest in the powerful cognitive computing tools that are ANNs can be explained by several factors. Based on the seminal work of \textcite{Rosenblatt}, it is known since the Universal Approximation Theorem of \textcite{Cybenko} that ANNs can approximated any continuous and finite function.
Data-driven algorithms, given a set of inputs and a set of outputs, they are able to estimate the mapping from the firsts to the seconds as long as they are provided with enough data and such a relationship exists. Thus, compared to econometric models, ANNs have the advantage of requiring almost no assumptions about the data generating process and of theoretically being able to model much more involved dynamics. In addition, some ANNs such as the long short-term memory neural networks (LSTMs) of \textcite{LSTM} display remarkable ability in detecting both short-term and long-term dependencies in time series, a useful feature for modeling asset return volatility and covariances that both exhibit high persistence over time. Also, thanks to their architecture inspired from the human brain, ANNs are able to sort out the information they are given in input in order to extract its most important features. This characteristic makes them able to process inputs of very large dimensions. However, when the number of input features is large, they tend to severely overfit in-sample and underfit out-of-sample if they are not provided with enough observations. As such, ANNs seem unlikely to directly overcome the curse of dimensionality faced by econometric models. 

While there have already been a number of papers trying to predict the volatility of one asset using neural networks, the study of their application to the case of the conditional covariance matrix of asset returns remains almost non-existent. To our knowledge, only \textcite{Caietal} and \textcite{Bucci} have addressed this issue so far. \textcite{Caietal} relied on the use of a Conditionally Restricted Boltzmann Machine (CRBM) to predict the Cholesky factorization of the one-step-ahead conditional covariance matrix for a sample of forty assets, and found that their model produced similar results to those of a CCC. \textcite{Bucci} also relied on the use of a Cholesky decomposition to ensure the positive definiteness of the predicted matrix, but with different types of ANNs and on a sample of only three assets. He found in particular that nonlinear autoregressive networks with exogenous inputs (NARX) and LSTMs allowed for a slight increase in precision compared to a DCC. While these two papers obtained rather encouraging results, they measured the accuracy of the models they proposed in a way that can be considered inappropriate. Indeed, both relied on the use of loss functions measuring the precision of the matrix forecasts element by element, an unsuitable choice in this framework according to \textcite{EngleColacito2006}. Also, due to the known sensibility of neural network to the curse of dimensionality, their models could not possibly be applied in a framework where the number of assets is in the hundreds.
As such, we believe that there is a need to propose a novel approach in covariance matrix forecasting using neural networks.

Because ANNs are sensitive to the ratio of the number of input features on the number of observations, rather than trying to predict the whole conditional covariance matrix, it seems more realistic to combine them with multivariate GARCHs that decompose volatility and correlation predictions such as the CCC or the DCC. Indeed, several papers have already shown that ANNs have a strong potential in forecasting univariate volatility and often outperform econometric models. As such, producing volatility forecasts with ANNs rather than with univariate GARCHs in a DCC might lead to an improved performance. In particular, a stream of literature has studied the ability of hybrid models, mixing univariate GARCHs and neural networks (GARCH-ANNs), to deliver better volatility predictions than purely econometric models. Despite presenting very promising results, the usefulness of such models in the multivariate case has yet to be investigated. Overall, two types of hybrid GARCH-ANNs can be found in the literature. The first, studied by \textcite{Donaldson} and \textcite{Bildirici} among others, uses the outputs of an ANN as exogenous variables within a univariate GARCH model. Although these papers found that such parametrizations allowed for a better performance than econometric models or simple ANNs, the second type introduced by \textcite{Roh} seems more flexible. As ANNs have the ability to process remarkably well inputs of various types, and because GARCH parameters contain useful information about the conditional volatility process on the data studied, \textcite{Roh} proposed to use GARCH parameters as inputs in ANNs. By testing his model on the KOSPI200 index, he found that it gave rise to a significant improvement in performance compared to non-hybrid ANNs. Subsequently, \textcite{Hajizadeh} and \textcite{Kristjanpoller} extended this idea by using asymmetric GARCH parameters as inputs. By testing their methodologies on the Istanbul stock exchange and three different Latin American markets respectively, both papers noticed an improvement over non-hybrid ANNs and econometric models. One drawback of these papers, however, is that they are based on the use of feedforward neural networks that are not well-suited for time series analysis, and thus ignore the potential of recurrent neural networks specially designed for sequential problems. \textcite{KimWon} recently addressed this by proposing the use of an LSTM to forecast the volatility of the KOSPI200 index. In addition to showing that LSTMs are more accurate than feedforward neural networks or econometric models for predicting volatility, they also found that adding the parameters of several types of univariate GARCHs as inputs at the same time improved model performance. Until now only used in the univariate case, these hybrid models and in particular the one of \textcite{KimWon} leave room for interesting possibilities in the multivariate case. 
\vspace{3mm}

The goal of this paper is thus to investigate whether or not multivariate hybrid models can deliver superior performance compared to state-of-the art multivariate GARCHs. To do so, we compare the various models in a Minimum Variance Portfolio framework \parencite{Markowitz}, which has the advantage of being entirely based on covariance predictions and is a common choice in the literature on the subject (see \cite{Engle2017}, \cite{Nakagawa}, \cite{DeNard}, \cite{Trucios}).

We contribute to the existing literature by proposing a new class of models that aims at taking advantage of the proven reliability of GARCH-ANNs in the univariate case in order to improve performance of traditional econometric models decomposing volatility and correlation predictions. To our knowledge, this is the first attempt at using hybrid models to predict the volatility of a large number of assets at the same time. In addition, the model proposed has the advantage of being built in a way that its performance could actually increase with the number of assets under consideration.

The remainder of this paper is organized as follows. In section \ref{sec:2} we recall the framework of the Minimum Variance Portfolio. In section \ref{sec:3} we describe the GARCH model of \textcite{garch}, as well as the DCC model and recent methods used to render the DCC robust in high-dimensions. In section \ref{sec:4} we describe the neural network architectures used in our model. In section \ref{sec:5}, we present the new hybrid model proposed. Finally, in section \ref{sec:6} the model is backtested using real-life data with a sample of 400 stocks.

%% file: MVP/MVP.tex
Throughout this paper, we consider a sample of $N$ assets with the price of asset $i \in \{1, ..., N\}$ at time $t$ denoted by $P_{i,t}$ and its (log) return by $y_{i,t}=\log(P_{i,t}/P_{i,t-1})$. We note $Y_t=(y_{1,t},...,y_{N,t}) \in \mathbb{R}^N$ the vector containing the daily returns of all the assets at time $t$ and $w_t=(w_{1,t},...,w_{N,t}) \in \mathbb{R}^N$ the weighting scheme of any portfolio constructed on our sample of $N$ assets at time $t$. Concretely, $w_{i,t}$ represents the percentage of the portfolio allocated to asset $i$, such that the daily return $y_{p,t}$ of a given portfolio $p$ with weighting scheme $w_t^p$ is given by $y_{p,t}=\ ({w_t^p})' Y_t$, where $({w_t^p})' \in \mathcal{M}_{1, N}(\mathbb{R})$\footnote{By $\mathcal{M}_{p,q}(\mathbb{R})$, we denote the space of matrices of dimension $p\times q$ with real-valued coefficients.} denotes the transpose of $w_t^p$.

As explained in the introduction, an advantage of the Minimum Variance Portfolio (MVP), described by \textcite{Markowitz}, is that its construction only relies on finding the weights which allow to minimize the conditional variance of the portfolio $\sigma_{p,t}^2=Var(y_{p,t}|\mathcal{F}_{t-1})$, where $(\mathcal{F}_{t})_{t\geq0}$ is a filtration on a given probability space $(\Omega, \mathcal{A}, \mathbb{P})$, such that $\mathcal{F}_{t-1}$ represents all the information available at $t-1$. Hence, the MVP without short selling is solution to the following minimization problem:

\begin{align}
    & \min_{w_t}\sigma_{p,t}^2= w_t' H_t w_t\\\
    & s.t.\  w_t' \mathbb{I}=1\\\
    & w_{i,t}\geq 0, i=1,...,N \label{eq:short-sell constraint}
\end{align}

\noindent
where $\mathbb{I} \in \mathbb{R}^N$ is a vector of ones. The constraint $w_t^\prime\mathbb{I}=1$ is often called the “feasibility constraint” and implies that at all time $t$, the portfolio must be fully invested. The no short-selling constraint (equation \ref{eq:short-sell constraint}) is sometimes imposed to portfolio managers due to the difficulty of respecting negative portfolio weights in reality.

In a minimum variance portfolio framework, the variable of interest is $H_t$ which represents here the true one-step-ahead conditional covariance matrix of asset returns, given by
\begin{equation}
    H_t=Var(Y_t|\mathcal{F}_{t-1}).
\end{equation}
It is an $N\times N$ symmetric matrix whose coefficient on the $i$-th row, $j$-th column is given by $\sigma_{ij,t}=\ Cov(y_{i,t},y_{j,t}|\mathcal{F}_{t-1})$, the conditional covariance between the returns of assets $i$ and $j$, its $i$-th diagonal coefficient is given by $\sigma_{i,t}^2=Var(y_{i,t}|\mathcal{F}_{t-1})$ the conditional variance of asset $i$ return. By definition, $H_t$ is a positive definite matrix, meaning that it is symmetric and that for all nonzero $N$-dimensional vector $x$ we must have $x^\prime H_tx>0$. 

\begin{remark}
    In practice, to avoid paying too much transaction costs, portfolios are generally rebalanced at most every month. Taking the example of a porfolio manager that uses monthly rebalancing, he will mainly be interested in predicting the monthly covariance matrix that we note $\Sigma_{\tilde{t}}$, where $\tilde{t}$ denotes the time period from day $t$ to day $t+20$ following the convention of 21 trading days per month. Using daily log returns this matrix is defined as 
\begin{equation}
    \Sigma_{\tilde{t}}=Var(Y_{t+20}+...+Y_t|\mathcal{F}_{t-1}),
\end{equation}
such that the optimal weights for a monthly rebalanced portfolio when short-selling is allowed are given by
\begin{equation}
    w_{\tilde{t}}^\ast= \frac{\Sigma_{\tilde{t}}^{-1}\mathbb{I}}{\mathbb{I}^\prime \Sigma_{\tilde{t}}^{-1}\mathbb{I}}. \label{eq:weights monthly MVP}
\end{equation}
\end{remark}
However, for the sake of simplicity, in this paper we will only proceed to produce one-step-ahead forecasts, meaning that we will consider the forecast of the covariance matrix for the first day of the following month to represent the monthly covariance matrix.


%% file: Econometric_models/GARCH.tex
Linear econometric models such as the linear regression or Autoregressive Moving Average (ARMA) processes are not well suited to model the conditional volatility of asset returns. As they rest on the assumption of homoskedastic errors, they assume that the conditional variance of the process is constant which goes against the observed volatility clustering. In order to overcome this limit, \textcite{arch} introduced the Autoregressive Conditional Heteroskedasticity (ARCH) model. By considering asset return conditional variance as heteroskedastic and modelizing it as a function of past volatility shocks, the ARCH was a real breakthrough allowing for a proper representation of volatility clustering. This class of models was further extended by \textcite{garch} who introduced the Generalized ARCH (GARCH).

    Let $p,q \in \mathbb{N}\backslash \{0\}$. We say that a stochastic process $(y_t)_{0\leq t\leq T}$, defined on $(\Omega, \mathcal{A}, \mathbb{P})$, is a univariate GARCH(p,q) process if it is in the form
\begin{align}
    & y_t= \mu_t+ \varepsilon_t\\\
    & \varepsilon_t= \sigma_t z_t\\\
    & \sigma_t^2= \omega+ \sum_{i=1}^{p}\alpha_i\varepsilon_{t-i}^2+ \sum_{i=1}^{q}\beta_i\sigma_{t-i}^2, \label{eq: GARCH(1,1)}
\end{align}
\noindent
where $y_t$ is the return of the asset at time $t$, $\mu_t$ is the conditional mean of $y_t$ such that $\mu_t=\ \mathbb{E}\left[y_t|\mathcal{F}_{t-1}\right]$. $z_t\overset{i.i.d}{\sim} \mathcal{N}(0, 1)$\footnote{$z_t$ can follow another distribution as long as it is a white noise} is a white noise such that the residual process $(\varepsilon_t)_{0\leq t \leq T}$ is characterized by the conditional moments $\mathbb{E}\left[\varepsilon_t|\mathcal{F}_{t-1}\right]=\mathbb{E}\left[\varepsilon_t\right]=0$ and $Var(\varepsilon_t|\mathcal{F}_{t-1})=\ \sigma_t^2$, where $\sigma_t^2=\ Var(y_t|\mathcal{F}_{t-1})$ is the one-step-ahead conditional variance of asset returns.

We see that the conditional variance, in a GARCH process, is dependant on both the intensity of past volatility shocks and past conditional volatilities. The difference between this model and the ARCH is that in the second there is no lagged conditional variance in the process. This simple addition from Bollerslev allowed to greatly reduce the number of lags needed for the model to deliver satisfying performances. In fact, as shown by \textcite{garchcomparison}, a GARCH(1,1) is generally sufficient to propose an efficient modeling of conditional variances whereas an ARCH(1) shows poor performance.

Note that in a GARCH(1,1)\footnote{for a higher-order GARCH, the restrictions are more involved and are not detailed here}, in order to obtain consistent forecasts it is necessary to impose that  $\omega \geq0$, $\alpha \geq0$, $\beta\geq0$ and $\alpha+\beta<1$. The positivity constraint ensures that the conditional variance predicted by the model is positive at all time whereas $\alpha+ \beta<1$ gives variance stationary errors, which allows for the conditional volatility to be mean-reverting. Indeed, in a GARCH(1,1), the unconditional variance of the residuals is given by
\begin{equation}
    Var\left(\varepsilon_t\right)=\ \mathbb{E}\left[\sigma_t^2\right]={(1-\ \alpha-\beta)}^{-1}\omega,
\end{equation}
which is defined if and only if $\alpha+\beta<1$. 

\begin{remark}
    Note that the GARCH model has some limits. Indeed, in a GARCH process, after a volatility shock, the conditional variance reverts to its unconditional mean at an exponential rate which is not consistent with the empirically observed hyperbolic decay of volatility shocks \parencite{timedepvol}. Also, as $z_t$ is typically assumed to follow a symmetric probability distribution, the GARCH model does not reflect the asymmetries in volatility movement that tend to differ depending on whether returns are negative or positive. To account for these empirical observations several models were introduced later on the as the Exponential GARCH \parencite{egarch} or the Fractionally Integrated GARCH \parencite{figarch}, but we do not detail these models here as they are not of primary importance to our work.
\end{remark}

%% file: Econometric_models/DCC.tex
As mentioned in the introduction, multivariate GARCH models that propose sophisticated parametrizations of future conditional covariance matrices typically suffer from the 'curse of dimensionality', as the number of parameters to estimate explodes with the number of assets considered. 

Several models were introduced by academics to overcome this dimensionality problem while keeping a realistic specification of the conditional covariance matrix. We focus here on the scalar Dynamic Conditional Correlation (DCC) GARCH introduced by \textcite{DCC}, which can be considered as one of the most (if not the most) popular models in the recent literature. 

Before detailing the Dynamic Conditional Correlation GARCH of \textcite{DCC}, we briefly introduce some key points about multivariate GARCH models. Let $(Y_t)_{t\geq 0}$ denote the return process with values in $\mathbb{R}^N$, and $(e_t)_{t\geq 0}$ the residual process with values in $\mathbb{R}^N$, both defined on the probability space $(\Omega, \mathcal{A}, \mathbb{P})$. In a multivariate GARCH model, they are defined as:
\begin{align}
    & Y_t= M_t+e_t\\\
    & e_t= H_t^{1/2}Z_t \label{eq:residuals MGARCH}
\end{align}
\noindent
where $M_t= \mathbb{E}[Y_t|\mathcal{F}_{t-1}] \in \mathbb{R}^N$ is the conditional mean vector of $Y_t$ and is often assumed to be null for daily returns. $Z_t \in \mathbb{R}^N$ is a white noise vector such that $\mathbb{E}\left[Z_t\right]=0$
and $Var(Z_t)= I_N$, with $I_N$ the identity matrix of order $N$. Then, $e_t \in \mathbb{R}^N$ the vector of errors (residuals) is characterized by the conditional moments $\mathbb{E}\left[e_t|\mathcal{F}_{t-1}\right]=0$ and $Var(e_t|\mathcal{F}_{t-1})= H_t$, where $H_t \in \mathcal{M}_N (\mathbb{R})$ is the one-step-ahead conditional covariance matrix of the return vector such that $Var(Y_t|\mathcal{F}_{t-1})= H_t$. Note that $H_t^{1/2}$ in equation \ref{eq:residuals MGARCH} is not unique but that any matrix verifying $H_t^{1/2}\left(H_t^{1/2}\right)^T= H_t$ works.

\subsubsection{The scalar DCC}

The main idea of the scalar DCC is to decompose covariance forecasts in volatility and correlation forecasts separately, so as to limit as much as possible the number of parameters to estimate while keeping realistic dynamics of the conditional covariance matrix.

The scalar DCC is defined as:
\begin{align}
    & Y_t |\mathcal{F}_{t-1}\sim \mathcal{N}(0,H_t)\\\
    & Y_t=e_t=D_t s_t\\\
    & H_t=D_t R_t D_t,
\end{align}
\noindent
where
\begin{equation}
    R_t=Diag(Q_t)^{-1/2}  Q_t  Diag(Q_t)^{-1/2}, \label{eq:adjustRDCC}
\end{equation}
\noindent
with
\begin{equation}
    Q_t=(1-\alpha-\beta)\bar{R} +\alpha s_{t-1} s_{t-1}'+\beta Q_{t-1},
    \label{eq:corr-targ-DCC}
\end{equation}
\noindent
where $D_t\in \mathcal{M}_N(\mathbb{R})$ is a diagonal matrix that contains the conditional volatility of each asset at time $t$, $R_t=Corr(Y_t |\mathcal{F}_{t-1})\in \mathcal{M}_N(\mathbb{R})$ is the conditional correlation matrix of asset returns. By definition, $R_t$ is a positive semidefinite matrix with only ones on its diagonal and all off-diagonal elements contained in the interval $[-1,1]$. $s_t = D_t^{-1} e_t \in \mathbb{R}^N$ denotes the standardized residuals at time $t$. These standardized residuals are a key feature of the DCC because of the fact that we have $Var(s_t |\mathcal{F}_{t-1})=Corr(Y_t |\mathcal{F}_{t-1})$, which allows for the modelization of $Q_t \in \mathcal{M}_N(\mathbb{R})$ using equation \ref{eq:corr-targ-DCC} that corresponds to a scalar BEKK multivariate GARCH \parencite{BEKKpaper} with correlation targeting \footnote{correlation targeting is an adaptation of the variance targeting approach of \textcite{vartarg}}. Note that for this correlation targeting to be usable it is necessary to ensure that $\alpha$ and $\beta$ are positive scalars and that $\alpha+\beta<1$, so that the unconditional correlation matrix of asset returns $\bar{R} = Corr(Y_t)\in \mathcal{M}_N(\mathbb{R})$, in equation \ref{eq:corr-targ-DCC}, is defined. The problem here is that the one-step-ahead conditional correlation matrix estimated through a multivariate GARCH is not ensured to be a true correlation matrix with only ones on its diagonal. This is why the matrix $Q_t$ that follows the scalar BEKK process is called a conditional “pseudo-correlation matrix”. In order to obtain a true correlation matrix, $R_t$ is then computed as $R_t=Diag(Q_t)^{-1/2}  Q_t  Diag(Q_t)^{-1/2}$ (equation \ref{eq:adjustRDCC}).

\begin{remark}
 Note that here we made the assumption that $Y_t |\mathcal{F}_{t-1}\sim \mathcal{N}(0,H_t)$ which is not mandatory, but seems to be the most common choice in the recent literature (\cite{Engle2017}, \cite{Nakagawa}, \cite{Trucios}).
\end{remark}

%% file: Econometric_models/Robustifying_multivariate_garch.tex
\paragraph{Standard Estimation Procedure of the DCC}

The 'classic' estimation procedure of the scalar DCC can be divided into three steps.
\vspace{1.5mm}

In a first step the residual process $(e_t)_t$, or the return process $(Y_t)_t$ in the case we assumed that $\mathbb{E}[Y_t|\mathcal{F}_{t-1}]=0$, is “DE-GARCHED” \parencite{Engle2009}. In order to DE-GARCH the series, $N$ univariate GARCHs are run (one for each asset) to compute the in-sample residuals, and the conditional variances that give us the in-sample matrix of conditional volatilities $D_t$ for each time step. Once we have obtained these matrices, we can compute the vector of standardized residuals $s_t=D_t^{-1} e_t$ for each time step, hence the term DE-GARCHING.

In a second step, the unconditional correlation matrix $\Bar{R}$ is estimated using a proxy, typically the sample correlation matrix
\begin{equation}
    \hat{\bar{R}}=T^{-1} \sum_{t=1}^{T}s_t s_t',
\end{equation}
where $T$ is the number of past observations available, which is an unbiased estimator of $\Bar{R}$ and is equal to the sample covariance matrix of the standardized residuals.

In a third step, the parameters $\alpha$ and $\beta$ of equation \ref{eq:corr-targ-DCC} are computed using their maximum likelihood estimators.
\vspace{1.5mm}

While the three steps procedure detailed above works well in small dimensions, when the number of assets increases it suffers from severe estimation difficulties and often leads to poor out-of-sample performance. There are two main points causing these issues. In the two following subsections we briefly present the methods used in the recent literature to address these.

\paragraph{Estimating the Unconditional Covariance Matrix: Nonlinear Shrinkage}

As we saw in the preceding subsections, variance targeting and correlation targeting are very useful in order to decrease the number of parameters to estimate. Those methods are based on the use of the unconditional covariance matrix $\Gamma \in \mathcal{M}_N(\mathbb{R})$ ($\Gamma$ for the returns and $\bar{R}$ for the standardized residuals), which as we have seen is generally estimated by its sample counterpart $\hat{\Gamma}$. This sample covariance matrix has the advantage of being an unbiased estimator that does not require complicated computations. However, as it relies on the estimation of $O(N^2)$ variables with $O(NT)$ data points, if the ratio $N/T$ is not small enough it can become a very noisy estimator. In fact, \textcite{Linshrink} show that, unless the ratio is negligible (for example less than $1/100$, see \cite{Engle2017}), the sample covariance matrix systematically overestimates the large eigenvalues of the unconditional covariance matrix and underestimates the small eigenvalues. This bias in eigenvalues is the result of an in-sample overfitting which leads to a poor out-of-sample performance of the sample covariance matrix. Thus, if $T$ is not large enough compared to $N$, multivariate GARCHs relying on variance or correlation targeting are very likely to underperform out-of-sample, leading to errors in portfolio optimization. The problem is that on financial markets, when the number of assets considered is large ($N=500$ for example), it is sometimes impossible to get enough observations per asset for the ratio $N/T$ to become negligible.

As the poor performance of the sample covariance matrix is linked to its biased eigenvalues, a way to provide a better performing estimator of the unconditional covariance matrix is to reduce this bias. This is the idea behind the method known as shrinkage. As the sample covariance matrix $\hat{\Gamma}$ is symmetric, according to the spectral theorem there exists a diagonal matrix $\Lambda = \text{diag}(\lambda_1, \ ..., \ \lambda_N) \in \mathcal{M}_N(\mathbb{R})$, where $(\lambda_i)_{i=1,...,N}$ are the sample eigenvalues, and an orthogonal matrix $U \in \mathcal{M}_N(\mathbb{R})$ whose columns are sample eigenvectors $(u_1, \ ..., \ u_N)$, where $\forall i \in \{1, ..., N\}$, $u_i$ is an eigenvector associated with the eigenvalue $\lambda_i$, such that
\begin{equation}
    \hat{\Gamma} =U\Lambda U'.
\end{equation}

\textcite{Linshrink} propose a linear shrinkage formula that pushes the eigenvalues of $\hat{\Gamma}$ towards their mean with an intensity determined by a parameter, such that the linear shrinkage estimator is optimal under general asymptotics, i.e. when $N$ and $T$ are both allowed to grow towards infinity but their ratio remained bounded. While in some cases their method allow for a significant improvement in the estimation of the unconditional covariance matrix, they note that in some cases, typically when the sample eigenvalues are too far away from each other, the method does not bring much improvement.

\textcite{Nonlinshrink} explain that it is due to the fact that in reality the problem is a nonlinear one. As such, they introduce a nonlinear shrinkage formula. To do so, they define a function called the \textit{Quantized Eigenvalues Sampling Transform} (QuEST). We do not detail the construction of this function here as it goes beyond the scope of this paper, but the general idea is as follows. Let $N, T \in \mathbb{N}$, that respectively represent the dimension of the covariance matrix and the number of observations available. Given a set of data, the QuEST function is a deterministic function 
\begin{align}
    Q_{T, N}: [0, +\infty)^N & \longrightarrow [0, +\infty)^N \\\
    x=(x_1, ..., x_N) & \longmapsto Q_{T,N}(x)=(q_{T,N}^1(x), ..., q_{T,N}^N(x))
\end{align}
that maps the population eigenvalues into the sample eigenvalues. They then use this function to approximate numerically the population eigenvalues by minimizing the quadratic function $x \in  [0, +\infty)^N \longmapsto \Vert Q_{T,N}(x) - \lambda \Vert_2^2$, where $\lambda = (\lambda_1, ..., \lambda_N) \in \mathbb{R}^N$, and $\Vert \cdot \Vert_2$ is the $l^2$ norm in $\mathbb{R}^N$, such that the approximation $\hat{\tau}=(\hat{\tau}_1, ..., \hat{\tau}_N) \in \mathbb{R}^N$ of the population eigenvalues is given by 
\begin{equation}
    \hat{\tau} = \text{argmin}_{x\in [0, +\infty)^N}\Vert Q_{T,N}(x) - \lambda \Vert_2^2.
\end{equation}
However, because it is impossible to approximate the population eigenvectors associated with these eigenvalues, using $\hat{\tau}$ would not be optimal. As such, \textcite{Nonlinshrink} then use a nonlinear shrinkage formula to compute the shrunk eigenvalues $(\hat{\lambda}(\hat{\tau})_i)_{i=1,...,N}$ that are optimal under general asymptotics. Without giving much detail about this nonlinear shrinkage formula, the idea is that $\hat{\tau}$ allows to approximate discretely, $\forall t \in \mathbb{R}$ where it is defined, the limit 
\begin{equation}
    \Tilde{m}_F(t) = \lim_{z \in \mathbb{H} \rightarrow t} m_F(z),
\end{equation}
where $\mathbb{H}=\{z \in \mathbb{C}; \ \text{Im}(z) > 0\}$, and $m_F$ is the Stieltjes transform of $F$ the limiting spectral distribution of the sample spectral distribution $F_T$. Then, $\forall i \in \{1, ..., N\}$, the shrunk eigenvalue $\hat{\lambda}(\hat{\tau})_i$ is computed as 
\begin{equation}
    \hat{\lambda}(\hat{\tau})_i = \frac{\lambda_i}{|1 - \frac{N}{T} - \frac{N}{T}\lambda_i \Tilde{m}(\lambda_i)|^2},
\end{equation}
where $\Tilde{m}$ is the numerical approximation of $\Tilde{m}_F$.
Finally, by defining $\hat{\Lambda}=\text{diag}(\hat{\lambda}(\hat{\tau})_1, ..., \hat{\lambda}(\hat{\tau})_N) \in \mathcal{M}_N(\mathbb{R})$, the shrunk sample covariance matrix $\hat{\Gamma}_{NLS}$ is given by
\begin{equation}
    \hat{\Gamma}_{NLS} = U\hat{\Lambda}U'.
\end{equation}

\paragraph{Estimating the Parameters: The Composite Likelihood Method}

In univariate and multivariate GARCHs, the parameters are typically estimated through a Maximum Likelihood Estimation (MLE). While this method allows to find consistent estimates in univariate GARCHs, it is problematic in the multivariate case when the dimension is large, even for the scalar DCC where there are only two scalar parameters to estimate.

 Assuming that the return process $(Y_t)_t$ is conditionally distributed as a multivariate Gaussian with zero mean such that $Y_t |\mathcal{F}_{t-1}\sim \mathcal{N}(0,H_t)$, its conditional density is given by:
\begin{equation}
    f(Y_t |\mathcal{F}_{t-1})=\frac{1}{2\pi^{N/2}\sqrt{|H_t|}} \exp\left(-\frac{1}{2}Y_t' H_t Y_t \right),
\end{equation}
\noindent
where $|H_t |$ is the determinant of the conditional covariance matrix at time $t$. Ignoring the constant term, the corresponding log-likelihood function is thus given by:
\begin{equation}
    l(\theta)=-\frac{1}{2} \sum_{t=1}^{T}\left[\log⁡(|H_t(\theta)|)+Y_t' H_t^{-1} (\theta)Y_t \right], 
\end{equation}
\noindent
with $\theta=(\alpha,\beta) \in \Theta$ the vector that contains the two parameters of the scalar DCC with correlation targeting (equation \ref{eq:corr-targ-DCC}), and $\Theta = \{(x, y) \in \mathbb{R}^2; \ x>0, \ y>0, \ x+y < 1\}$ the set of possible parameters.

Like for the univariate GARCH, there is no closed-form solution and the MLE requires to rely on a numerical optimization. The problem, as highlighted by \textcite{Pakel} among others, is that the optimization requires to invert the conditional covariance matrix for each time step until convergence. As the computational load of inverting an $N\times N$ matrix is $O(N^3)$, the numerical optimizer is very slow when considering large-dimensional matrices. Furthermore, \textcite{Pakel} show, using Monte-Carlo simulations, that the MLE gives a more and more biased estimate of $\alpha$ as $N$ increases with $T$ constant, and that when $T\leq N$ the estimator fails to converge rendering the use of MLE impossible in that case.

To overcome this issue, they propose to rely on a Composite Likelihood (CL) that approaches the full-dimensional likelihood through a combination of bivariate likelihoods. Pairwise composite likelihood works as follows. The first step is to select a number $P$ of pairs of assets from the sample of $N$ assets. \textcite{Pakel} test two methods. Either all the possible combinations of pairs are picked, i.e. $O(N^2)$ pairs, or only contiguous pairs are picked, i.e. $O(N)$ pairs. Then, for each pair note $Y_{i,t}=(y_{i_1,t},y_{i_2,t})\in \mathbb{R}^2$ the $2\times 1$ vector containing the returns at time $t$ of the two assets in pair $i$ with conditional covariance matrix $H_{i,t} \in \mathcal{M}_2(\mathbb{R})$. By considering the vector of parameters $\theta$ to be constant among all pairs, the log-likelihood function of pair $i$ is then given by:
\begin{equation}
    l_{i}(\theta)=-\frac{1}{2} \sum_{t=1}^{T}\left[\log⁡(|H_{i,t}(\theta)|)+Y_{i,t}' H_{i,t}^{-1} (\theta)Y_{i,t} \right].
\end{equation}
\noindent
The total composite (log) likelihood function is then defined as
\begin{equation}
    l_{CL}(\theta)=\frac{1}{P} \sum_{i=1}^{P}l_{i}(\theta),
\end{equation}
\noindent
which allows to obtain $\hat{\theta}=(\hat{\alpha},\hat{\beta})$ a consistent estimator of $\theta$ such that $\hat{\theta}=\text{argmax}_{\theta \in \Theta}{l_{CL}(\theta)}$.

As it only requires to invert $2\times 2$ matrices, the Maximum CL Estimation (MCLE) has a computational cost of $O(P)$ (i.e. $O(N)$ or $O(N^2)$ depending on the selection of pairs), which represents an important gain compared to the MLE. In practice (see for example \cite{Engle2017}, \cite{Nakagawa}, \cite{Trucios}) it is the MCLE based on contiguous pairs which is implemented. The reason behind this is that \textcite{Pakel} found it to be approximately 240 times faster than the one based on all combinations of pairs for $N=480$ while yielding very similar performances. \textcite{Pakel} note that
even though MCLE loses some efficiency compared to MLE as it only observes bivariate likelihoods, it displays much better performances when the number of assets is large. The parameters obtained are not subject to the same increase in bias as in the MLE, and the MCLE converges even when $T\leq N$.

%% file: Deep_Learning/Intro_section.tex
As mentioned in the introduction, artificial neural networks (ANNs) have shown promising performance for volatility forecasts. Their ability to modelize very elaborate functions as well as the existence of architectures specially designed for sequential problems make them an appealing alternative to traditional econometric models. In this section we detail three different architectures of ANNs that we use in our hybrid model.



%% file: Deep_Learning/FNN.tex
Deep Feedforward Neural Network (DFNs) are the “representative” ANNs in the sense that other ANNs are variants based on their architecture. DFNs are data-driven parametric function approximators. Given a set of inputs $\mathcal{X}$ and a set of outputs $\mathcal{Y}$, they approximate the (unknown) function $\varphi:x\in \mathcal{X}\longmapsto y \in \mathcal{Y}$, that maps the inputs into the outputs, by a parametric counterpart $\hat{\varphi}(\ \cdot\ ; \theta): x \in \mathcal{X} \longmapsto \hat{\varphi}(x; \theta)$, where $\theta \in \Theta$, with $\Theta$ some parameter set. The 'form' of the function $\hat{\varphi}(\cdot\ ; \theta)$ and the dimension of the parameter set $\Theta$ are determined by the architecture of the network, which is controlled by a set of predetermined hyperparameters.





\begin{figure}
    \centering
    \includegraphics[width=130mm, height=85mm]{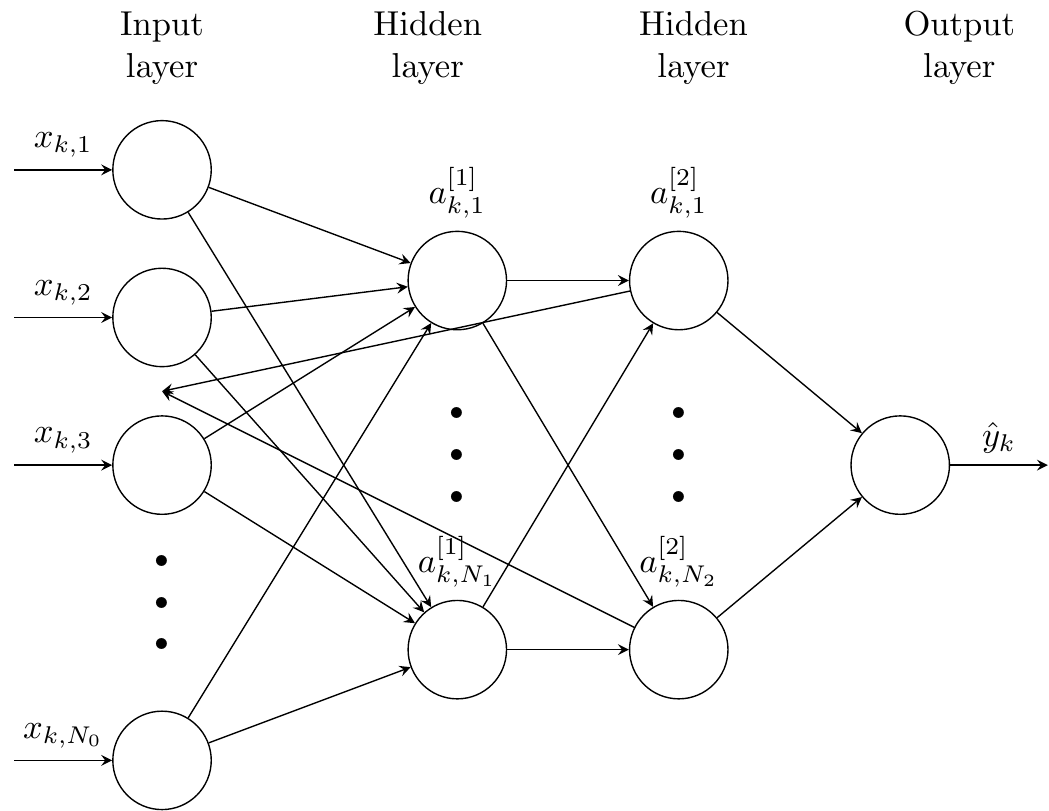}
    \caption{Simple dense feedforward neural network with two hidden layers}
    \label{fig:dfn}
\end{figure}

Concretely, DFNs are based on a layered structure (see Figure \ref{fig:dfn} for a visual example). They possess an input layer, one or several hidden layers and one output layer, each composed of several nodes that are generally called neurons. To understand how these universal approximators work, let us first introduce a few notations. Let $L+1$ the number of layers and $N_l,\ l=0,\ \ldots,\ L$ the number of nodes in the $l$-th layer such that the input layer is of dimension $N_0$ and the output layer of dimension $N_L$. Let $K$ the number of training examples $\left(x_k,\ y_k\right),\ k=1,\ \ldots,\ K$, with $x_k$ the input for the $k$-th training example and $y_k$ the targeted output.

In DFNs, the information only goes forward. The output estimation $\hat{y}_k$ of the $k$-th training example is computed through a sequential process among the layers. Starting from the input layer, in the $k$-th training example, information flows into the first layer in the form
\begin{equation}
    z_k^{\left[1\right]}=W^{\left[1\right]}x_k+\ b^{\left[1\right]},
\end{equation}
where $z_k^{\left[1\right]}\in \mathbb{R}^{N_1}$ represents the information entering the first layer, $b^{\left[1\right]} \in \mathbb{R}^{N_1}$ is a vector of biases, and $W^{\left[1\right]} \in \mathcal{M}_{N_1, N_0}(\mathbb{R})$ is a matrix of weights that determines the relative importance of each input feature. The vector $z_k^{\left[1\right]}$ is then 'activated' using an activation function $f^{\left[1\right]}: \mathbb{R}^{N_1} \rightarrow \mathbb{R}^{N_1}$, which is typically a differentiable nonlinear function, such that the value output by the first layer in the $k$-th training example is 
\begin{equation}
    a_k^{\left[1\right]}=f^{\left[1\right]}\left(z_k^{\left[1\right]}\right).
\end{equation}

This iterative process then continues for the second layer which outputs an $N_2$ dimensional vector $a_k^{\left[2\right]}$ computed as
\begin{equation}
    a_k^{\left[2\right]}=f^{\left[2\right]}\left(z_k^{\left[2\right]}\right) = f^{\left[2\right]}\left(W^{\left[2\right]}a_k^{\left[1\right]}+\ b^{\left[2\right]}\right).
\end{equation}
The sequential process continues until the output layer is reached where the estimation $\hat{y}_k$ of the output for the $k$-th training example is computed as 
\begin{equation}
    {\hat{y}}_k=f^{\left[L\right]}\left(W^{\left[L\right]}f^{\left[L-1\right]}\left(\ldots f^{\left[2\right]}\left(W^{\left[2\right]}f^{\left[1\right]}\left(W^{\left[1\right]}x_k+\ b^{\left[1\right]}\right)+b^{\left[2\right]}\right)\ldots\right)+b^{\left[L\right]}\right),
    \label{eq:DFN}
\end{equation}
with $f^{\left[l\right]}: \mathbb{R}^{N_l} \rightarrow \mathbb{R}^{N_l} $ the activation function of the $l$-th hidden layer, $W^{\left[l\right]} \in \mathcal{M}_{N_l, N_{l-1}}(\mathbb{R})$ the matrix of weights determining the importance of the activated values of the layer $l-1$ for the features of the layer $l$ and $b^{\left[l\right]} \in \mathbb{R}^{N_l}$ the vector of biases for the $l$-th layer.

The matrices of weights and the vectors of biases, for each layer, are the parameters of the network, which put together give the parameter $\theta$ of the beginning of the section. They are computed by the DFN through training such that the distance between the set of estimated outputs $\hat{y}$ and the set of true outputs $y$, in the training sample, is minimum according to a pre-defined loss function. This training is typically done through Gradient descent and Backpropagation algorithms which require the activation functions to be differentiable.

Several academics have compared the ability of simple or hybrid DFNs and univariate GARCHs to produce asset return volatility forecasts (see \cite{Bildirici}, \cite{Hajizadeh}, \cite{KimWon} for example). While the results were generally in favor of the DFNs, such neural networks are in fact not well-suited for time series modeling. Because in a DFN the information only goes forward from the input to the outputs without any feedback connection, they are unable to detect sequential patterns in the data. To circumvent this limit for sequential problems, new ANNs called Recurrent Neural Networks were introduced.


%% file: Deep_Learning/RNNs.tex
Recurrent Neural Networks (RNNs) were introduced by \textcite{RNNs}. They possess special layers, called recurrent layers, that allow them to process sequential data more efficiently than DFNs. The specificity of these layers is that, instead of directly propagating the information forward to the next layer like hidden layers in a DFN do, they can store information and reuse it to analyze other input vectors.

\begin{figure}[!htb]
    \centering
    \includegraphics[width=110mm, height=90mm]{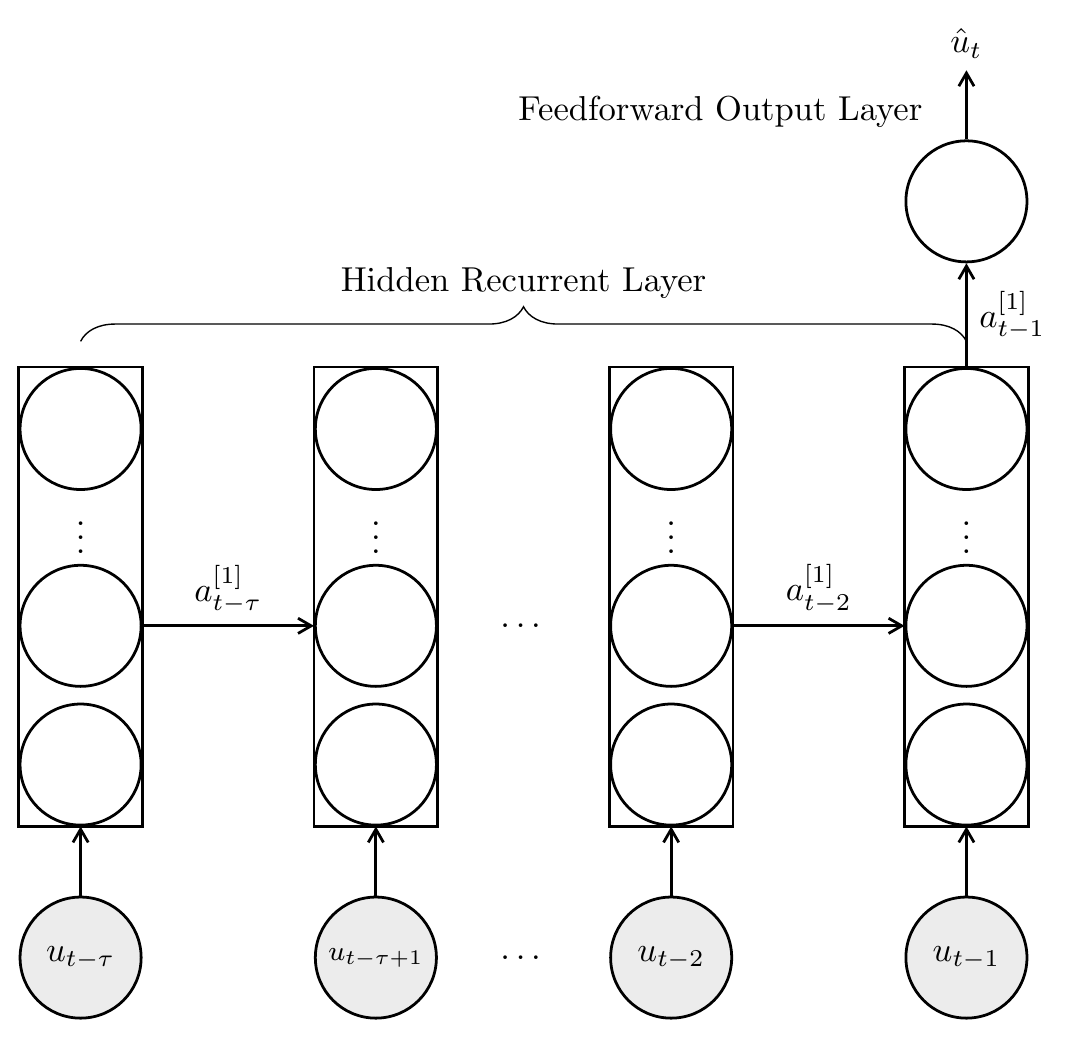}
    \caption{Simple RNN with one hidden recurrent layer and one feedforward output layer.}
    \label{fig:RNN}
\end{figure}

On Figure \ref{fig:RNN} is presented a visual example of a very simple RNN, composed of only one hidden recurrent layer and a feedforward output layer. All the rectangles represent the same recurrent layer composed of $N_1$ neurons, drawn several times to show how it processes temporal information.


As the RNN takes as inputs a sequence of vectors and not an isolated vector like a DFN does, 
the $k$-th training example $(x_k, y_k)$ can be defined as $x_k=\left\{u_{t-\tau}, \ ..., u_{t-1} \right\}$, $y_k = u_t$, where $\tau \in \mathbb{N}^*$, and $(u_t)_{0\leq t \leq T}$ is the process that we observe, such that the model uses the last $\tau$ observations of this process to predict its value at time $t$.

To find temporal patterns in time series, RNNs treat the sequential inputs in a chronological order. Starting from the first vector of the input sequence in the $k$-th training example, the recurrent layer extracts information in the form
\begin{equation}
\label{eqn:info1stvect}
    z_{k,t-\tau}^{\left[1\right]}=W_xu_{t-\tau}+b^{\left[1\right]},
\end{equation}
with $z_{k,t-\tau}^{\left[1\right]}\in \mathbb{R}^{N_1}$, $W_x \in \mathcal{M}_{N_1, N_0}(\mathbb{R})$ a matrix of weights determining the importance of each feature of the input vector considered and $b^{\left[1\right]}\in \mathbb{R}^{N_1}$ a vector of biases. Similarly to a DFN, the RNN transforms this information into a $N_1$ vector of activated values 
\begin{equation}
    a_{k,t-\tau}^{\left[1\right]}=g^{\left[1\right]}\left(z_{k,t-\tau}^{\left[1\right]}\right)
\end{equation}
with $g^{\left[1\right]}$ a nonlinear differentiable activation function. The difference compared to a DFN is that, instead of directly passing this transformed information to the following layer, the recurrent layer stores it in its memory. This stored information is called the ``hidden state" of the recurrent layer. Then, when the neural net goes on to analyze the second element of the sequence, it uses as an input not only $u_{t-\tau+1}$ but also $a_{k,t-\tau}^{\left[1\right]}$. The two are combined such that the new flow of information entering the layer is in the form
\begin{equation}
    z_{k,t-\tau+1}^{\left[1\right]}=W_xu_{t-\tau+1}+W_a a_{k,t-\tau}^{\left[1\right]}+b^{\left[1\right]},
\end{equation}
where $W_x$ and $b^{\left[1\right]}$ are identical to the ones in equation \ref{eqn:info1stvect} and $W_a \in \mathcal{M}_{N_1, N_1}(\mathbb{R})$ is a matrix that weights the importance of the hidden state features. After activation, $a_{k,t-\tau+1}^{\left[1\right]}=g^{\left[1\right]}\left(z_{k,t-\tau+1}^{\left[1\right]}\right)$ becomes the new hidden state.
After analyzing all the vectors of the input sequence, the recurrent layer outputs a vector of activated values that uses all the information extracted from the preceding vectors in the form
\begin{equation}
\label{eqn:last output recurrent layer}
    a_{k,t-1}^{\left[1\right]}=g^{\left[1\right]}\left(z_{k,t-1}^{\left[1\right]}\right)=g^{\left[1\right]}\left(W_xu_{t-1}+W_ag^{\left[1\right]}\left(\ldots g^{\left[1\right]}\left(W_xu_{t-\tau}+b^{\left[1\right]}\right)\ldots\right)+b^{\left[1\right]}\right).
\end{equation}

The ability of RNNs to approximate involved nonlinear functions combined with their predisposition for time series problems make them very appealing in our framework. However, classic RNNs such as the one presented here suffer from a major pitfall as most of the time, they fail to capture long term patterns in the data due to the problem known as ``vanishing gradient". \textcite{BPTTHoch91}, \textcite{Bengio1994} and \textcite{Pascanu2013} provide a detailed explanation of this phenomenon

%% file: Deep_Learning/LSTM.tex
Long Short-Term Memory (LSTM) neural networks are a special type of RNN introduced by \textcite{LSTM} to model efficiently both short-term and long-term dependencies in the data. The general architecture of a LSTM network is exactly the same as the one of classic RNNs but their recurrent layers are composed of different units. Instead of possessing layers of neurons, LSTM networks are characterized by memory blocks. Those memory blocks consist of memory cells that can store the information for a very long time and of gates that allow to decide which information should be kept and which should be forgotten by the network.

Consider the same specification of the training examples as in the Section \ref{sec:3.1.2} just above. The information processing of the element $u_{t-i}$, of the input sequence, in an LSTM layer is characterized by the following equations:
\begin{align}
    & {\widetilde{c}}_{t-i}=\tanh{\left(W_{ac}a_{t-i-1}+W_{xc}u_{t-i}+b_c\right)}  \label{eq:c tilde}\\\ 
    & G_{u,\ t-i}=\text{sig}\left(W_{au}a_{t-i-1}+W_{xu}u_{t-i}+b_u\right)  \label{eq:update gate}\\\ 
    & G_{f,\ t-i}=\text{sig}\left(W_{af}a_{t-i-1}+W_{xf}u_{t-i}+b_f\right)  \label{eq:forget gate}\\\ 
    & G_{o,\ t-i}=\text{sig}\left(W_{ao}a_{t-i-1}+W_{xo}u_{t-i}+b_o\right)  \label{eq:output gate}\\\ 
    & c_{t-i}=G_{u,\ t-i}\circ{\widetilde{c}}_{t-i}+G_{f,\ t-i}\circ\ c_{t-i-1} 
    \label{eqn: memory cell}\\\
    & a_{t-i}=G_{o,\ t-i}\circ\tanh{\left(c_{t-i}\right)}. \label{eq:hidden state LSTM}
\end{align}
Note that in order to lighten the notations, we dropped the subscript $k$ but equations \ref{eq:c tilde} to \ref{eq:hidden state LSTM} describe the processing of one vector of the input sequence in one training example.

Here, $c_{t-i} \in \mathbb{R}^{N_1}$ is a vector that represents the state of the memory cell after analyzing $u_{t-i}$. $G_{u,\ t-i}$, $G_{f,\ t-i}$ and $G_{o,\ t-i}$ are all $N_1$-dimensional vectors that represent respectively the update gate, the forget gate and the output gate. ${\widetilde{c}}_{t-i}$, sometimes called the input modulation gate, can be seen as a candidate to replace $c_{t-i-1}$ in the memory cell. All the matrices noted $W$ are matrices of weights whose coefficients differ according to the computation they are used for, but are shared among all vectors of the input sequence like in an RNN. $\text{sig}: \ x\in \mathbb{R} \mapsto \frac{1}{1+e^{-x}}$ and $\tanh: \ x\in \mathbb{R} \mapsto \frac{e^x-e^{-x}}{e^x+e^{-x}}$ are the sigmoid and hyperbolic tangent functions, two of the most common activation functions in ANNs. They have values contained in the intervals $]0,1[$ and $]-1,1[$ respectively.

\begin{figure}
    \centering
    \includegraphics[width=120mm, height=85mm]{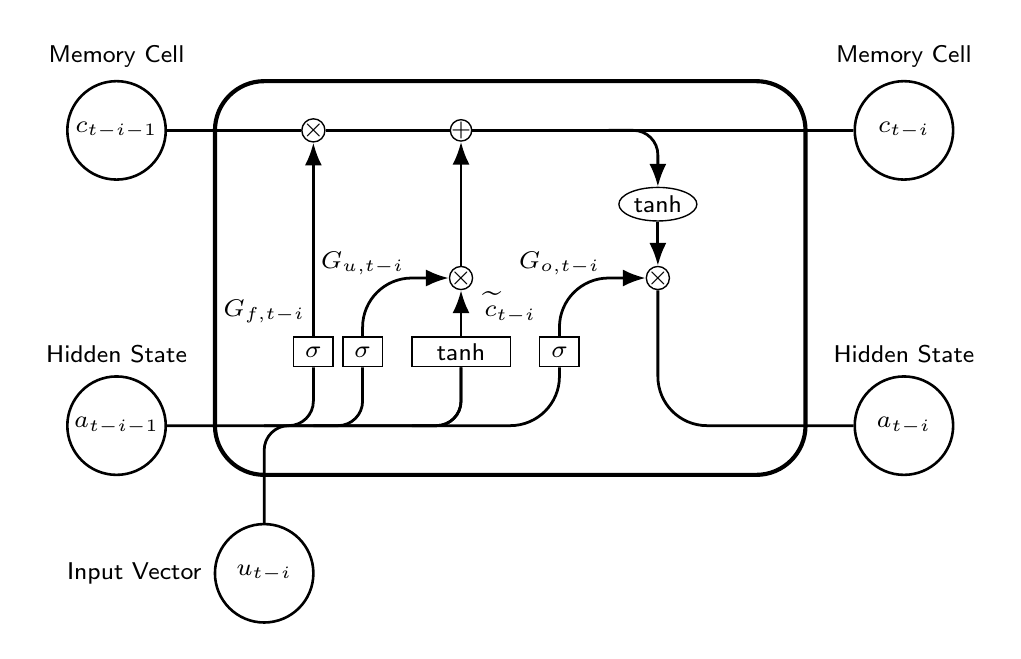}
    \caption{Memory block representing a LSTM layer where the variables are $N_1$ dimensional vectors. In reality, the LSTM layer is composed of $N_1$ blocks like this one that output scalars.}
    \label{fig:LSTM cell}
\end{figure}

To understand how the LSTM layer processes the input vector $u_{t-i}$, a simplified LSTM layer is presented on Figure \ref{fig:LSTM cell}. 
The LSTM layer starts by extracting from its hidden state $a_{t-i-1}$ and the input vector $u_{t-i}$ the information needed to compute the input modulation gate (${\widetilde{c}}_{t-i}$), the update gate ($G_{u,\ t-i}$), the forget gate ($G_{f,\ t-i}$) and the output gate ($G_{o,\ t-i}$). The states of these four gates are computed as a weighted sum (transformed through activation) of $a_{t-i-1}$ and $u_{t-i}$, using different sets of parameters for each gate (equations \ref{eq:c tilde}, \ref{eq:update gate}, \ref{eq:forget gate}, \ref{eq:output gate}). 
The update, forget and input modulation gates are then used to calculate the new state of the memory cell $c_{t-i}$, as shown in equation \ref{eqn: memory cell}. The purpose of this cell is to allow the LSTM to retain over long periods of time the information useful to estimate the outputs. At each time step, the memory cell is computed as a combination of the information $c_{t-i-1}$ it contained at the previous time step and the new information contained in ${\widetilde{c}}_{t-i}$. The forget gate determines which elements of $c_{t-i-1}$ should be retained or forgotten by the LSTM at the new time step and the update gate determines which elements of ${\widetilde{c}}_{t-i}$ are of interest and should be added to the memory cell. Using Hadamard product in equation \ref{eqn: memory cell} allows to consider individually the utility of each element of $c_{t-i-1}$ and ${\widetilde{c}}_{t-i}$. As $G_{u,\ t-i}$ and $G_{f,\ t-i}$ are computed using a sigmoid function applied element-wise, all their coefficients are between $0$ and $1$. As such, the more the LSTM layer judges that an element of  $c_{t-i-1}$ or ${\widetilde{c}}_{t-i}$ is useful, the closer the value of the corresponding element in the corresponding gate will be to $1$. On the contrary, if the LSTM judges that an element is not worth keeping in memory, the value in the corresponding gate will be close to $0$. Note that initially, there was no forget gate in the LSTM of \textcite{LSTM}, which led to a constantly increasing state for the memory cell ($c_{t-i}$). It was \textcite{Gersetal.} who introduced the forget gate in order for the LSTM to be able to drop preceding information considered unimportant.

Once the memory cell has been updated, it is activated using the $\tanh$ function. Then, the new hidden state of the LSTM layer is computed using the output gate and the activated memory cell, as shown in equation \ref{eq:hidden state LSTM}. Like the forget and update gates, the output gate $G_{o,\ t-i}$ only has values between $0$ and $1$, which allows the network to determine which elements of $\tanh{{(c}_{t-i})}$ are useful to calculate the new hidden state of the layer.
\vspace{3mm}

To see the math behind the ability of LSTMs to overcome the vanishing gradient problem, one can refer to the original paper of \textcite{LSTM}. The general idea here is just that through the use of its gates and memory cells, the LSTM is able to sort out the information extracted from input vectors and  to remember it throughout the whole input sequence or to forget it when necessary. Thanks to these mechanisms, LSTMs can efficiently identify both long-term and short-term dependencies in the data. In that sense, given the empirically observed strong persistence over time of conditional volatilities and correlations, they seem particularly well suited to produce forecasts of the conditional covariance matrix compared to other ANNs. Various researchers have already tested the ability of LSTMs to produce volatility forecasts such as \textcite{GoogleLSTM}, \textcite{KimWon}, \textcite{Zhou2018} and found that LSTMs (hybrid or not) outperformed both DFNs and the traditional GARCH(1,1). 
For these reasons, we choose to use LSTMs in the hybrid models explained in the following section.

%% file: Hybrid_model/Intro_section_hybrid.tex
As explained in the introduction, previous studies by \textcite{Roh}, \textcite{Hajizadeh} and \textcite{Kristjanpoller} tested hybrid models, using the parameters of GARCH-type processes as inputs in DFNs, to produce forecasts of asset return conditional volatility. More recently, \textcite{KimWon} used the same methodology but with an LSTM, which seems more appropriate considering the strength of this network for time series predictions. Also, they found that adding the parameters of several GARCH models at the same time helped to improve the accuracy of the neural network.
In all those papers, hybrid models showed superior performances compared to simple ANNs or GARCH models.

The performance of these hybrid models can be explained by the ability of ANNs to efficiently process various types of inputs at the same time. This ability is due to their structure that allows them to sort the information contained in the data and to extract from it features that help improving predictions. 
In that sense, the use of variables extracted from GARCH-type models as additional inputs is beneficial since they provide indications about the empirical characteristics of the conditional volatility process. Indeed, unlike ANNs that are ``black-boxes" in which the features of the hidden layers cannot really be interpreted by humans,
in each of the various GARCH models the parameters represent pre-defined features that explain variations in conditional volatility over time.

In a GARCH(1,1) for example, the term $\beta \sigma_{t-1}^2$ in equation \ref{eq: GARCH(1,1)} represents the persistence of asset return conditional volatility over time, i.e. what is the effect of the conditional variance of an asset return at time $t-1$ on its conditional variance at time $t$.
The term $\alpha \varepsilon_{t-1}^2$, on the other hand, can be interpreted as representing the impact of yesterday’s volatility shock on the current volatility level.

\begin{remark}
    Note that while \textcite{KimWon} showed that adding parameters from other types of GARCH models, in addition to those of a GARCH(1,1), as inputs in the neural net increased the performance, here we only use a GARCH(1,1) for the sake of simplicity. However, it would be rather easy to add parameters from other types of GARCH models in the hybrid model we propose.
\end{remark}

In what follows, we propose a way to generalize hybrid GARCH-ANNs to the case of several assets that combines volatility forecasts made by neural networks with correlations predicted by econometric models, to see if it can improve the performance of models like the CCC or the DCC.

%% file: Hybrid_model/The_GARCH-LSTM-DCC.tex
Recall the DCC specification for the one-step-ahead conditional covariance matrix:
\begin{equation*}
    H_t=D_t R_t D_t,
\end{equation*}
with $D_t$ the matrix of conditional volatilities and $R_t$ the matrix of conditional correlations. The GARCH-LSTM-DCC is simply a hybrid DCC in which the elements of $D_t$ are forecast using a GARCH-LSTM on conditional volatilities. Based on the observation that hybrid models mixing GARCHs and neural networks seem to produce better performances than univariate GARCHs, the motivation behind this model is to try to enhance purely econometric DCCs. 

 To predict the volatility of $N$ assets, it would have been possible to use as inputs hybrid vectors containing the volatilities of all the assets, and to use as the output the vector containing their volatilities at the following time step. However, the problem with such specification is that while neural networks are able to efficiently process inputs of very large dimensions and can work well with millions of parameters, they are sensitive to the ratio of input features over training examples. If the number of observations is too small compared to the number of elements in the input vectors, ANNs tend to overfit the data on the training sample which generally leads to poor performance out-of-sample. If we used this specification, we would have a number of input and output features on the order $O(N)$, which would far too high for the neural network to produce accurate forecasts given only $T$ observations. For these reasons, we choose here to integrate within the same LSTM all the assets in the sample, but to construct the input vectors such that the neural network considers only one asset at a time. 
 
 More precisely, the GARCH-LSTM-DCC model is constructed as follows. We start by fitting the selected GARCH-type models on the training sample for each asset. This allows to extract the quantities of interest that will be integrated as inputs in the neural net, and to compute at the same time the vectors of standardized residuals $(s_t)_{0\leq t \leq T}$ that will be used to forecast the conditional correlation matrix. Then, for all assets and for all time steps in the training sample, we compute the ex-post estimators of asset return daily volatility. This gives us for each asset $i=1,...,N$ a vector $d_i=(d_{i,1},\ \ldots,\ d_{i,T})'\in \mathbb{R}^T$, with $T$ the number of time steps in the training sample and $d_{i,t}$ the ex-post estimator of asset i true volatility on day $t\in \{1, ..., T\}$. Those volatilities are then combined into $T$ different hybrid vectors per asset, meaning that we have $NT$ hybrid vectors in total. Let $X_{i,t}$ the hybrid vector of asset $i$ at time $t$. This vector is of dimension $1+m$, with $m$ the number of GARCH parameters extracted per asset. Because we only use a GARCH(1,1) here, we have $X_{i,t}=(d_{i,t},\ \alpha_i\varepsilon_{i,t}^2,\ \beta_i\sigma_{i,t}^2)'\in \mathbb{R}^3$, where $\alpha_i\varepsilon_{i,t}^2$ and $\beta_i\sigma_{i,t}^2$ are the GARCH features obtained for asset $i$ at time step $t$.

The LSTM is then trained on all the assets at the same time which means that it uses the same matrices of weights and the same biases for each asset. In that sense, the specification proposed here might not be as flexible as using an input vector containing the volatilities of all the assets. However, thanks to the GARCH parameters used as inputs, the dynamics of the volatility of each asset will be different despite the identical parameters used by the network. In addition, we propose in section \ref{sec:5.3} a very simple way to add flexibility to volatility dynamics in the model.

Let $\tau \in \mathbb{N}\backslash \{0\}$ the lag order of the LSTM. In the GARCH-LSTM-DCC model we have a total of $K:= T - \tau$ training examples per asset. Meaning that we have a total of $NK$ training examples and $1+m$ features per asset. To simplify, let $\left(x_{i,k},y_{i,k}\right)$, $i=1,\ldots,N$, $k=1,\ldots,K$ the $k$-th training example for asset $i$, we define it as
\begin{align}
    & x_{i,k}=\left\{X_{i,t-\tau},\ldots,X_{i,t-1}\right\} \\\
    & y_{i,k}=d_{i,t},
\end{align}
such that the LSTM learns for each asset how to estimate the one-step-ahead conditional volatility (more precisely its ex-post estimator) using the last $\tau$ hybrid vectors of this asset. The out-of-sample forecast of the one-step-ahead conditional covariance matrix is then given by
\begin{equation}
    {\hat{H}}_t={\hat{D}}_t{\hat{R}}_t{\hat{D}}_t,
\end{equation}
where ${\hat{D}}_t$ is estimated using the neural net and ${\hat{R}}_t$ using an econometric model such as the DCC.

%% file: Hybrid_model/Advantages_and_limits.tex
In addition to the potential superior forecasting power of univariate GARCH-LSTMs compared to traditional univariate GARCHs, the GARCH-LSTM-DCC model has the advantage that the number of training examples scales linearly with $N$. In that sense, it differs from the traditional DCC where the quality of the conditional volatility forecasts do not depend on the number of assets considered. Here, the more assets we include in our sample, the larger the training set grows and the more the neural network can learn precisely how to model the volatility dynamics.

Nonetheless, it possesses some limits. As all assets are processed together, it is unsure whether or not the model will be able to show good performance, and we cannot know beforehand if using the LSTM has some advantage over simply using $N$ univariate GARCHs to model conditional volatilities. Also, since $R_t$ is predicted using an econometric model, it is subject to the curse of dimensionality described in the introduction and can only be estimated using an overly simplistic specification. As such, compared to multivariate GARCH models, the GARCH-LSTM-DCC only allows at best to potentially increase the accuracy of conditional volatility predictions, but does not overcome the trade-off between flexibility and computational feasibility.

To circumvent the potential lack of flexibility of the GARCH-LSTM-DCC, in which the neural network applies the same weights to the volatility of each asset, we simply propose to use one-hot-encoding. One-hot-encoding is a basic Machine Learning technique that consists in adding to the input for each training example of the asset $i=1, \ldots, N$, a vector $h_i = (h_{i, 1}, \ldots, h_{i, N})' \in \mathbb{R}^N$, where for $j=1, \ldots, N$,
\begin{equation*}
    h_{i, j} = 0 \text{ if } i\neq j \text{, and } h_{i, j} = 1 \text{ if } i=j.
\end{equation*}
Adding the one-hot-encoded input to the network allows the neural network to differentiate the assets and to apply specific weights to each, thus introducing more flexibility in the GARCH-LSTM-DCC model.

\begin{remark}
    In the GARCH-LSTM-DCC implemented in section \ref{sec:6}, the one-hot-encoded input was not added directly to the LSTM layer, but was concatenated with the LSTM outputs before serving as an input to the following feedforward layers.
\end{remark}

\begin{remark}
    Note that using only GARCH(1,1) inputs, it is relatively easy to produce multi-steps-ahead forecasts with the GARCH-LSTM-DCC by proceeding in an iterative fashion. The only technical point when doing so is to compute the out-of-sample residuals of the GARCH(1,1). However, as equation \ref{eq: GARCH(1,1)} provides us with a closed-form formula for these, doing so does not present any difficulty.
\end{remark}

%% file: Empirical_part/Intro_section_empirical.tex
In this section, we compare the performance of the GARCH-LSTM-DCC, with and without GARCH inputs and with and without one-hot-encoding, to the one of the DCC. To do so, we measure the practical usefulness of these models for asset allocation decisions by constructing the minimum variance portfolio using real-world stock data. Details about the numerical implementation and the architecture of the neural networks are provided in appendix \ref{app:Numerical Details}.

%% file: Empirical_part/Data.tex
The dataset used in this thesis was provided by Clevergence.  It is composed of the daily adjusted close prices from 12/02/2004 to 06/11/2020 of 700 stocks that belonged to the SP500 index somewhere between 2004 and 2020.

To simplify the analysis, only stocks with data available from the beginning to the end of the observed period were used in the implementation. Indeed, it would otherwise have been impossible to train the neural networks on certain stocks and including them in the out-of-sample analysis would then have led to distorted performance. In the end, we end up with 400 different stocks in our sample and 4023 daily observations for each or 4022 daily returns. Following the convention of 21 consecutive trading days for one month, we collected the first 1250 price observations in the training sample on which the models were trained, which gave an out-of-sample period of 2772 days or 132 months, from 09/29/2009 to 06/11/2020.

%% file: Empirical_part/Economic_Evaluation.tex
As written by \textcite{EngleColacito2006}, metrics such as the Frobenius norm that measure forecast accuracy element by element are not adapted to the problem of forecasting the covariance matrix of asset returns.  In reality, investors will be more interested in knowing whether or not a certain model can bring them added value compared to others, a thing not possible with only a statistical analysis of the performance. The main reason why typical loss functions are not adapted to the problem is because they give equal importance to each element of the covariance matrix, while in reality all elements do not have the same impact on the performance of the portfolio.

For the preceding reasons, in this paper we compare the performance of the different models through the use of the Minimum Variance Portfolio framework with short-selling allowed, which is a typical choice in the recent literature on the subject (see \cite{Engle2017}, \cite{Nakagawa}, \cite{DeNard}, \cite{Trucios}).

The portfolio generation rules are as follows. Over the 1250 days in-sample, from the 12/02/2004 to the 09/28/2009, the neural networks are trained to learn how to predict the targeted outputs using their respective inputs. Then, for the 132 months out-of-sample, we build 132 monthly portfolios for each model. The choice of a monthly rebalanced portfolio allows for more realistic results than if we were to rebalance on a daily basis, while allowing for more observations than if we were to select a quarterly or annual rebalancing. 

More specifically, on the last of the 21 days of each month, we predict the covariance matrix of the following day with all the models implemented. The weights of each portfolio are then rebalanced so as to minimize their variance computed with the matrix forecast. These weighting schemes are kept constant during the 21 days of the month, at the end of which the process is then repeated. Note that while the DCC model was trained every month, neural networks were only trained once per year to shorten the time needed for the backtesting of the model.
\vspace{2mm}

\noindent
To compare the performance of the hybrid models with the one of multivariate GARCHs, the following portfolios are constructed:
\begin{itemize}
    \item \textbf{1/N:} the equally weighted portfolio which is a common benchmark.
    
    \item \textbf{DCC:} model of \textcite{Engle2017}. It consists in a scalar DCC of \textcite{DCC}, in which nonlinear shrinkage of \textcite{Nonlinshrink} is applied to the sample correlation matrix $\hat{\bar{R}}$ used for correlation targeting in equation \ref{eq:corr-targ-DCC}.  Here, the forecast $\hat{D}_t$ is obtained using $N$ GARCH(1,1).
    
    \item \textbf{LSTM-DCC-OH:} model where the forecast $\hat{D}_t$ is obtained with a non-hybrid LSTM with one-hot-encoding, taking only the proxies of past conditional volatilities as inputs, and then combined with the matrix $\hat{R}_t$ predicted by the DCC above to obtain $\hat{H}_t$. 
    
    \item \textbf{G-LSTM-DCC-OH:} same as above but with GARCH(1,1) features as additional inputs in the neural network.
    
    \item \textbf{LSTM-DCC:} same as the LSTM-DCC-OH but without one-hot-encoding inputs.
    
    \item \textbf{G-LSTM-DCC:} same as the G-LSTM-DCC-OH but without one-hot-encoding inputs.
\end{itemize}

\begin{remark}
    For all the neural networks we used $d_{i,t} = |y_{i,t}|$ as the proxy for the volatility of asset $i\in \{1, \ldots, N\}$ at time $t$. Note that while this proxy is a common choice for daily volatility, using intraday returns to compute the proxy is likely to be more accurate and as such to lead to better results.
\end{remark}

After constructing the 6 portfolios detailed above, we obtain at the end of the out-of-sample period a sequence of daily returns for each. Let $Y^{\left(a\right)}=\{y_1^{\left(a\right)},\ldots,y_{2772}^{\left(a\right)}\}$ the 2772 daily returns observed out-of-sample for the model $a$. On each day $t \in \{1, \ldots, 2772\}$, the daily return of portfolio $a$ is computed as $y_t^{\left(a\right)}=(w_{\tilde{t}}^{\left(a\right)})\prime Y_t$, with $w_{\tilde{t}}^{\left(a\right)}$ the weighting scheme of the corresponding month and $Y_t$ the $N$-dimensional vector containing the returns of all the stocks considered on day $t$. 

Following \textcite{Engle2017} and \textcite{Nakagawa}, the performance of the portfolios is then compared through the use of three different out-of-sample measures. The first one is the annualized average return (\textbf{AV}) of the portfolio which is computed as 
\begin{equation}
    AV=252\bar{Y}^{\left(a\right)},
\end{equation}
with $\bar{Y}^{\left(a\right)}$ the empirical mean of the sequential returns for this portfolio. The second one is the annualized standard deviation (\textbf{SD}) such that
\begin{equation}
    SD=\sqrt{252}\sigma^{\left(a\right)},
\end{equation}
with $\sigma^{\left(a\right)}$ the standard deviation of $Y^{\left(a\right)}$. The third one is the information ration (\textbf{IR}), defined as 
\begin{equation}
    IR=\frac{AV}{SD}.
\end{equation}
Although having a high average return or a good information ratio is always desirable for an investor, following \textcite{Engle2017} we attach more importance to the annualized standard deviation. Indeed, being in a minimum variance portfolio framework, our primary goal is to minimize the risk and therefore to obtain the smallest possible SD, regardless of the AV or the IR.

%% file: Empirical_part/Results.tex
Table \ref{table:short} shows the annualized standard deviation (SD), the annualized average return (AV) and the information ratio (IR) for the 6 Minimum Variance Portfolios. For each metric, the best result is in \textcolor{blue}{\textbf{bold}}.
\vspace{3mm}

\input{Empirical_part/Table_results}

As explained in Section \ref{sec:6.1.2}, we consider the SD to be the most important performance indicator for the MVP, whose fundamental goal is to minimize it. We notice first that all models using neural networks outperform both the 1/N portfolio and the DCC in terms of SD. The G-LSTM-DCC-OH portfolio has the smallest annualized standard deviation, followed in order by the LSTM-DCC-OH, the G-LSTM-DCC, the LSTM-DCC and then the DCC. As such, we see that the neural network structure proposed in section \ref{sec:5} to forecast volatility of asset returns in high-dimensions is definitely of interest for market practitioners. Also, both the use of one-hot-encoding and GARCH features as inputs lead to a substantial decrease in annualized standard deviation and are thus recommended. Finally, it is notable that the best model, the G-LSTM-DCC-OH, outperforms the DCC by a wide margin, thus proving the interest of this model compared to the purely econometric ones. 

In terms of AV, we notice that only the G-LSTM-DCC-OH manages to beat the DCC. All the other models lead to significantly lower annualized average returns and struggle to beat the 1/N portfolio. This is reflected in the Information Ration of the portfolios. Indeed, for this metric only the G-LSTM-DCC-OH, once again, outperforms the DCC. All the other neural networks based models, although they are above the 1/N portfolio, underperform in terms of IR compared to the DCC.


%% file: Empirical_part/Table_results.tex
\begin{table}[ht]

\centering 

\begin{tabular}{c| C{1cm} C{1cm} C{1cm} } 
\hline
\multicolumn{4}{c}{09/29/2009 to 06/11/2020}\\
\hline 
Models & SD & AV & IR\\
\hline 
1/N & 19.16 & 8.13 & 0.48 \\ 
\hline
DCC & 13.74 & 10.13 & 0.74 \\
LSTM-DCC-OH & 12.32 & 8.29 & 0.67 \\
G-LSTM-DCC-OH & \textcolor{blue}{\textbf{12.04}} & \textcolor{blue}{\textbf{10.74}} & \textcolor{blue}{\textbf{0.89}} \\
LSTM-DCC & 13.26 & 7.55 & 0.57\\
G-LSTM-DCC & 12.66 & 8.69 & 0.68 \\
 [1ex] 
\hline 
\end{tabular}
\caption{Out-of-sample metrics on the whole period for minimum variance portfolios with short-selling allowed. The SD and AV refer to the annualized standard deviation and the annualized average return respectively. They are both expressed in percentages to ease the interpretation. For each metric the best result is in bold.} 
\label{table:short} 
\end{table}


%% file: Conclusion/Conclusion.tex
The goal of this paper was to investigate the ability of hybrid models, mixing GARCH processes and neural networks, to forecast covariance matrices of asset returns.

To do so, we presented the GARCH-LSTM-DCC model, a new model generalizing hybrid GARCH-ANNs models to the multivariate case. This model combines multi-dimensional volatility forecasts using LSTM neural networks, taking GARCH features as inputs, with correlation predictions using a multivariate GARCH model. We also investigated the interest of using GARCH features and one-hot-encoding on stock tickers as inputs in such model.

To determine the practical usefulness of the model proposed, we evaluated the out-of-sample performance of the minimum variance portfolios constructed using the predictions of this model with GARCH inputs or not and with one-hot-encoding or not. To do so, we used daily data from a sample of 400 stocks, that once belonged to the SP500 index, to predict one-step-ahead covariance matrices.

Our analysis led to the following conclusions. First, within the framework of the minimum variance portfolio, the addition of GARCH inputs and of the one-hot-encoding are both beneficial to the model in order to reduce the annualized standard deviation of the portfolio returns. This observation is consistent with previous studies on the univariate case that found that adding GARCH parameters as inputs in the neural networks helped to improve their predictions. Second, we found that the GARCH-LSTM-DCC produces particularly interesting performances. Indeed, when both types of inputs were added to the model, it significantly outperformed the DCC model both in terms of annualized standard deviation and in terms of annualized average return.

Furthermore, one should note that the performance of the GARCH-LSTM-DCC as presented here could be easily improved. Indeed, while the econometric models were trained every month, we only trained the neural networks once per year. By training them more often, the performance should logically increase even more. Also, we used a quite basic architecture for the neural network and did not bother trying to optimize its hyperparameters. By using methods such as Bayesian optimization, one could easily find a better set of hyperparameters. Moreover, we did not try to include other types of univariate GARCH models as input features, which is supposed to improve the accuracy of the model, at least in the univariate case. Finally, other types of layers are worth exploring, such as Convolutional layers which are getting popular in sequential learning problems when combined with LSTM layers.

To conclude, the GARCH-LSTM-DCC model seems to be of practical interest for market practitioners, in the sense that even with all these possible improvements unexplored it still managed to outperform by a wide margin its purely econometric counterpart, the DCC.

%% file: Acknowledgement/ack.tex
I would like to thank Thibaud Vienne, founder of Clevergence and professor of Machine Learning in the MSc 203 'Financial Markets' of Université Paris-Dauphine - PLS, for providing the dataset used in the empirical part, for giving me the opportunity to work on this subject, and for giving me the general direction of the research, i.e. producing forecasts of covariance matrices of asset returns using Deep Learning.

%% file: Appendices/Numerical_details.tex
The whole implementation of the model was done on Python. More specifically, we relied on Kaggle's GPU to run the neural networks. The neural networks were constructed using Keras functional API on tensorflow, which allows for a greater flexibility compared to the traditional Keras API (notably to merge the one-hot-encoding with the LSTM outputs). 

Regarding the implementation of the econometric models, we used the arch package of Kevin Sheppard to implement the GARCH(1,1) model. For the DCC, since there are no packages in Python, we developed it by ourselves by coding the composite likelihood method for parameters estimation and the necessary part to produce forecasts. For the writing of the composite likelihood part, we based ourselves on the code of the R package xdcclarge of Kei Nakagawa. 

Regarding the architecture of the neural networks, we first passed the hybrid inputs (without one-hot-encoding) inside a CuDNNLSTM layer with 100 neurons, a tanh activation function and a dropout of 0.4. Then, the output of this layer was concatenated with the one-hot-encoding (if present in the model). The concatenated vector was then passed into 4 Dense layers with respectively 350, 300, 250, 200 neurons each and a dropout of 0.5 each. All the dense layers used a sigmoid activation function. Finally, the result was passed through an output Dense layer with 1 neuron and a sigmoid activation function.